\documentclass[aps,prl,twocolumn,superscriptaddress,showpacs]{revtex4}

\usepackage{amsmath}    
\usepackage{amssymb}    
\usepackage{graphicx}   
\usepackage{epsf}
\usepackage{epsfig}
\begin{document}

\title{Zurek-Kibble Mechanism for the Spontaneous Vortex
Formation in $Nb-Al/Al_{ox}/Nb$ Josephson Tunnel Junctions: New
Theory and Experiment}
\thanks{Submitted to Phys. Rev. Letts.}

\author{R.\ Monaco}
\affiliation{Istituto di Cibernetica del C.N.R., 80078, Pozzuoli,
Italy and Unita' INFM-Dipartimento di Fisica, Universita' di
Salerno, 84081 Baronissi, Italy.}\email[Corresponding
author.e-mail:\ ] {roberto@sa.infn.it}
\author{J.\ Mygind}
\affiliation{Department of Physics, B309, Technical University of
Denmark, DK-2800 Lyngby, Denmark.} \email{myg@fysik.dtu.dk}
\author{M.\ Aaroe}
\affiliation{Department of Physics, B309, Technical University of
Denmark, DK-2800 Lyngby, Denmark.}\email{s001827@student.dtu.dk}
\author{R.\ J.\ Rivers}
\affiliation{Blackett Laboratory, Imperial College London, London
SW7 2AZ, U.K. }\email{r.rivers@imperial.ac.uk}
\author{V.\ P.\ Koshelets}
\affiliation{Institute of Radio Engineering and Electronics,
Russian Academy of Science, Mokhovaya 11, Bldg 7, 125009, Moscow,
Russia.}\email{valery@hitech.cplire.ru}
\date{\today}

\begin{abstract}
New scaling behavior has been both predicted and observed in the
spontaneous production of fluxons in quenched $Nb-Al/Al_{ox}/Nb$
annular Josephson tunnel junctions as a function of the quench time,
$\tau _{Q}$. The probability $f_{1}$ to trap a single defect during
the N-S phase transition clearly follows an allometric dependence on
$\tau _{Q}$ with a scaling exponent $\sigma = 0.5$, as predicted
from the Zurek-Kibble mechanism for {\it realistic} JTJs formed by
strongly coupled superconductors. This definitive experiment
replaces one reported by us earlier, in which an idealised model was
used that predicted $\sigma = 0.25$, commensurate with the then much
poorer data. Our experiment remains the only condensed matter
experiment to date to have measured a scaling exponent with any
reliability.

\end{abstract}

\maketitle

\medskip The Zurek-Kibble (ZK) scenario \cite{zurek1,zurek2,kibble1} for
continuous phase transitions proposes that transitions take effect
as fast as possible i.e. the domain structure initially reflects
the causal horizons. This proposal is amenable to direct testing
for transitions whose domain boundaries carry topological charge.
This is the case for Josephson Tunnel Junctions (JTJs), where the
topological charge is a fluxon i.e. a supercurrent vortex carrying
a single quantum of magnetic flux $\Phi_0 = h/(2e)$ in the plane
of the oxide layer between the two superconductors that make up
the JTJ. In the case considered in this paper of annular JTJs,
obtained by the superposition of two superconducting rings, it
corresponds to a magnetic flux that threads just one of the two
rings an odd number of times.

In 2000 an idealised model was proposed \cite{KMR&MRK} to test the
ZK scenario for JTJs, assuming that causal horizons are constrained
by the velocity of electromagnetic waves in the JTJ, the Swihart
velocity \cite{Swihart,Barone}. As a result, the probability $f_{1}$
for spontaneously producing one fluxon in the thermal quench of a
symmetric annular JTJ of circumference $C$ was predicted to scale
with the quench time $\tau _{Q}$ (the inverse quench rate) as

\begin{equation}
f_{1}\simeq \frac{C}{\bar{\xi}}=\frac{C}{\xi _{0}}\bigg(\frac{\tau _{Q}}{%
\tau _{0}}\bigg)^{-\sigma }.  \label{P1}
\end{equation}

\noindent In Eq.(\ref{P1}), ${\bar\xi}$ is the Zurek-Kibble causal
length, the correlation length of the relative phase angle at the
time of defect formation. It is defined through Eq.(\ref{P1}) in
terms of the cold correlation length $\xi_0$, the relaxation time
$\tau_0$ of the long wavelength modes and the quench time $\tau_Q$,
in turn defined by: $T_{C}/\tau _{Q}=-(dT/dt)_{T=T_{C}}$.
Eq.(\ref{P1}) holds for ${C<\bar\xi}$.

\noindent Under the assumptions of a) weak coupling of the
superconductors and b) exact critical slowing of the Swihart
velocity at the critical temperature $T=T_{c}$, we predicted $\sigma
= 0.25$ \cite{KMR&MRK}.

In 2001 our first proof-of-principle experiment \cite{Monaco1&2} was
performed, to test the scaling law of Eq.(\ref{P1}). The experiment
consisted of taking an annular JTJ isolated from its surroundings
and making it undergo a forced phase transition by heating it above
its superconducting critical temperature and letting it to cool
passively back towards the liquid He temperature. Once the thermal
cycle is over, the junction I-V curve is inspected and any trapped
fluxon can be detected by the appearance of  current peaks at
discrete voltages in the I-V characteristic of a JTJ. By detecting
the voltage position of these peaks the number of trapped fluxons
(or antifluxons) can be determined.

\noindent The experiment of Ref.\cite{Monaco1&2} was remarkably
successful, as the only experiment of the several \cite
{grenoble,helsinki,lancaster,lancaster2,technion,carmi,technion2,
Kirtley,pamplona,florence} previously performed to test scaling on
condensed matter systems which was sensitive enough to show genuine
ZK scaling behavior. However, this preliminary experiment suffered
from severe restrictions; in particular a limited range of cooling
rates, statistically poor data and above all insufficient shielding
of the earth magnetic field with the possibility of systematic
errors. The outcome was $\sigma$ commensurate with $0.25$.

In this paper we shall present results from a new experiment
designed to circumvent these problems, that we shall discuss later.
Our experiment shows extremely reliable scaling behavior of the form
(\ref{P1}), but with $\sigma = 0.5$ to high accuracy. This is
obviously at total variance with our earlier prediction. However, as
we shall show, the value $\sigma = 0.25$ depended critically on both
assumptions a) and b) cited earlier being {\it exactly} satisfied.
For realistic JTJs these assumptions are only {\it approximately}
satisfied in the immediate vicinity of $T=T_c$. Since defects form
very close to $T_c$, we shall see that this is sufficient for the
value of $\sigma$ to jump from $0.25$ to $\sigma = 0.5$ {\it
exactly}. Thus, rather than its negation, our experiment provides an
even more robust demonstration of the ZK scenario.

\medskip To reiterate, the theory in Ref.\cite{KMR&MRK} was developed for
JTJs whose electrodes are {\it weak coupling} superconductors; for
such JTJs the temperature dependence of the critical current density
$J_c(T)$ is given by the Ambegaokar-Baratoff equation \cite{AB}:

\begin{equation}
J_{c}(T)=\frac{\pi }{2}\frac{\Delta (T)}{e\rho _{N}}\tanh \frac{\Delta (T)}{%
2k_{B}T},  \label{JcA&B}
\end{equation}

\noindent where $\Delta (T)$ is the superconducting gap energy and
$\rho _{N}$ is JTJ normal resistance area squared.
Eq.(\ref{JcA&B}) provides a linear decrease of $J_c$ near $T_c$:
\begin{equation}
J_{c}(T(t))= \alpha J_{c}(0)\bigg(1-\frac{T}{T_{c}}\bigg)\simeq
\alpha J_c(0)\frac{t}{\tau_Q}, \label{JcTc}
\end{equation}

\noindent where $T(t=0) = T_c$ and the dimensionless quantity
$\alpha$ is approximately equal to $2 \Delta (0)/k_{B}T_{C}=3.5$.
However, our JTJs are based on $Nb$, a {\it strong-coupling}
superconductor, for which Eq.(\ref{JcA&B}) is not necessarily valid.
In practice, high quality and reproducible barriers are achieved by
depositing a thin $Al$ overlay onto the $Nb$ base electrode which
will be only partially oxidized, leaving a $Nb-Al$ bilayer
underneath having a non-BCS temperature dependence of the energy gap
and of the density of states . The proximity effect in
$Nb-Al/Al_{ox}/Nb$ JTJs has been extensively studied and it is known
to influence the electrical properties of the junctions, such as the
current-voltage characteristic and the temperature dependence of the
critical current density. Specifically, the proximity effect in
$S-N-I-S$ junctions is responsible for a subdominant temperature
dependence of the critical current density \cite{Rowell&Smith} in
the vicinity of $T_c$ of the form:

\begin{equation}
J_{c}(T(t)) \simeq \alpha' J_{c}(0) \left( 1-
\frac{T}{T_c}\right)^2 = \alpha' J_{c}(0)\left(\frac{t}{\tau
_Q}\right)^{2}, \label{JctProx}
\end{equation}

\noindent where $\alpha'$ is a constant depending intricately on the
degree of proximity. The last equation models the tail shaped
dependence of $J_c$ vs. $T$ near $T_c$; it has been theoretically
derived and experimentally confirmed by Golubov {\it et al.}
\cite{golubov} in 1995. Rephrasing the arguments of
Ref.\cite{KMR&MRK} with Eq.(\ref{JctProx}) replacing
Eq.(\ref{JcTc}), the Josephson penetration depth $\lambda_{J}(T)$,
which plays the role of system equilibrium coherence length
$\xi(T)$, near $T_c$ diverges linearly with time:

\begin{equation}
\xi(T(t))=\lambda _{J}(T(t))=\frac{\xi _{0} \tau _{Q}}{t},
\label{xeq1prox}
\end{equation}

\noindent where
\begin{equation}
\xi_{0}=\sqrt{\frac{\hbar }{2e\mu _{0}d_{s}\alpha' J_{c}(0)}},
\label{csi_not}
\end{equation}

\noindent $d_s$ being electrode thickness .

${\dot{\xi}}(t)=d\xi(t)/dt<0$ measures the rate at which the defects
contract, i.e., the speed of interfaces between ordered and
disordered ground states. Since ${\dot{\xi}}(t)$ decreases with time
$t>0$, the {\it earliest} possible time $t$ at which defects could
possibly appear is determined by causality,

\begin{equation}
{\dot{\xi}}(\bar{t})=-{\bar{c}}(\bar{t}),
\label{causality}
\end{equation}

\noindent where ${\bar{t}}$ is the causal time and $\bar{c}$ is
the Swihart velocity.

\noindent As we said, in Ref.\cite{KMR&MRK} we had assumed that the
Swihart velocity vanished at $T_c$ whereas, for realistic JTJs, it
just becomes very small. Swihart \cite{Swihart} has demonstrated
that for a thin-film superconducting strip transmission line the
solution for the velocity varies continuously  as one passes through
the critical temperature into the normal state. As a result, we
assume $\bar{c}(t)=\bar{c}_{nn}$ near the transition temperature
where $\bar{c}_{nn}$ is the speed of light in a microstrip line made
of normal metals. In the case of a microstrip line made by two
electrodes having the same thickness $d_{s}$ and the same skin depth
$\delta$, with $d_{s}<< \delta$, separated by a dielectric layer of
thickness $d_{ox}$ and dielectric constant $\epsilon$, $\bar{c}_{nn}
\simeq (2/\delta) \sqrt { d_{ox} d_{s}/ \epsilon \mu}$
\cite{Swihart}. Its value depends on the temperature very weakly,
but depends on the frequency $f$ through $\delta=\sqrt{ \rho / \pi
\mu f}$, $\rho$ being the normal metal residual resistivity.

\noindent  The match $\bar{c}(T_c)= \bar{c}_{nn}$ is certainly
realistic and we still have approximate critical slowing down
insofar as $\bar{c}_{nn}$ is much smaller than the zero temperature
Swihart velocity $\bar{c}_0 \simeq \sqrt {d_{ox}/{2 \lambda_{L0}
\epsilon \mu}}$, i.e., when the zero temperature London penetration
depth $\lambda_{L0} << \delta^2 / d_s$. For $300nm$ thick Nb
electrodes ($\rho= 3.8 \mu \Omega cm$ and $\lambda_{L0}=90nm$),
$\delta \simeq 1mm$ at say $f= 10 kHz$, so the last inequality is
fully satisfied. At the same frequency, for a value of the specific
barrier capacitance $c_s=\epsilon /d_{ox}=0.027 F/m^2$ typical of
low current density $Nb-Al/Al_{ox}/Nb$ JTJs, we get $\bar{c}_{nn}=6
\cdot 10^3 m/s$ and $\bar{c}_0= 1.6 \cdot 10^7 m/s$ \cite{note1}.

\noindent The solution of the causality equation
Eq.(\ref{causality}) with a non-vanishing Swihart velocity yields:
$\bar{t}= \sqrt{ {\xi_{0} \tau_{Q}/{ \bar{c}_{nn}}}}$.

\noindent Inserting the value of ${\bar{t}}$ into
Eq.(\ref{xeq1prox}) we obtain the new Zurek length $\bar{\xi}$:

\begin{equation}
{\bar{\xi}}= \xi({\bar{t}})=\sqrt{\xi_{0} \tau_{Q}  \bar{c}_{nn}}=\xi _{0}\bigg(\frac{\tau _{Q}}{\tau _{0}}%
\bigg)^{1/2}, \label{xiZprox}
\end{equation}

\noindent where $\tau _{0}=\xi _{0}/ \bar{c}_{nn}$ ($\tau
_{0}=O(1ns)$). We reach the important conclusion for realistic JTJs
that the probability $f_{1}$ for spontaneously producing one fluxon
in the quench is still predicted to scale with the quench time $\tau
_{Q}$ according to Eq.(\ref{P1}), but the critical exponent is now
$\sigma =\,0.5$, rather than  $\sigma =\,0.25$. Detailed
calculations will be reported elsewhere \cite{Next}.

\noindent However, it is worthy to observe that by varying $\tau_Q$
in the experimentally achieved four decade range $1 ms < \tau _{Q} <
10 s$, we get  $10 \mu s < \bar{t} < 1 ms$ that is always much
larger than $\tau_0$; it means that by the time the Josephson phase
\emph{freezes} the Josephson effect is well established. Further, in
the same $\tau_Q$ range the normalized freezing temperature
$\bar{T}/{T_C}$ at which the defect is formed is $0.99 <
\bar{T}/{T_C} <0.9999$. It would be really hard, if not impossible,
to measure the temperature dependence of $J_c$ and $\bar{c}$ so
close to $T_C$. We have then to resort to theoretical predictions.

\medskip The new experiment has a faster and more reliable single
heating system, obtained by integrating a meander line $50\mu$m
wide, $200$nm thick, and $8.3$mm long Mo resistive film in either
ends of the 4.2mm$\times$3mm$\times$0.35mm  Si chip containing the
Nb/AlOx/Nb trilayer JTJs. These resistive elements have a nominal
resistance of $50\Omega$ at LHe temperatures and, due to their good
adhesion with the substrate, are very effective in dissipating heat
in the chip. In fact, voltage pulses a few $\mu$s long and a few
volts high applied across the integrated heater provided quench
times as low as $1\mu$s, that is more than two orders of magnitude
smaller than for the previous situation \cite{Monaco1&2}. Further,
automatization of thermal cycles was implemented that allowed for
much more robust statistics to be achieved. At the end of each
thermal quench the junction I-V curve is automatically stored and an
algorithm has been developed for the detection of the trapped
fluxons. Finally, all the measurements have been carried out in a
magnetic and electromagnetically shielded environment. The low
frequency magnetic shielding was achieved by using $\mu$-metal,
cryoperm and lead cans. During the quench the JTJ was also
electrically isolated.

The experimental results reported here are restricted to one of two
annular JTJs with similar geometry to the sample used in the earlier
experiment ($C=500\mu$m and $d_s=300nm$), but with about 50 times
lower critical current density ($J_{c}(0)\simeq 60$A/cm$^2$): this
leads to a value of $\xi _{0}= 17 \mu m$ (we have assumed
$\alpha'=\alpha=3.5$ in  Eq.(\ref {csi_not})). An extensive
description of the chip layout, the experimental setup and the
quenching data will be given elsewhere \cite{Next} while the details
of the fabrication process can be found in Ref. \cite{VPK}. The
quench time $\tau _{Q}$ was continuously varied over more than four
orders of magnitude (from $20$s down to $1 ms$) by varying the
helium exchange gas pressure inside the vacuum can and/or the width
and the amplitude of the voltage pulse across the integrated
resistive element.

\begin{figure}[ht]
\begin{center}
\epsfysize=6.5cm \epsfbox{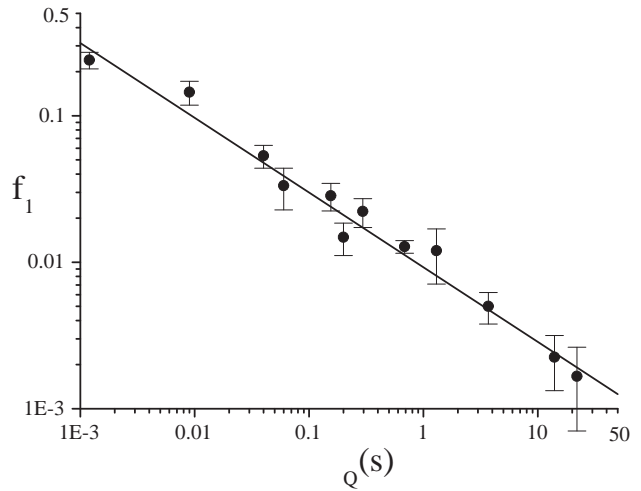}
\end{center}
\caption{Log-log plot of the measured frequency $f_{1}$ of trapping
single fluxons versus the quenching time $\tau _{Q}$. Each point
corresponds to many thermal cycles. The vertical error bars gives
the statistical error while the relative error bars in $\tau _{Q}$
amounting to $\pm10\%$ are as large as the dots' width. The solid
line is the fit to an allometric relationship $f_{1}=a\,\tau
_{Q}^{-b}$ which yields  $a=0.0092\pm 10\%$ (taking $\tau _{Q}$ in
seconds) and $b=0.51\pm 6\%$.} \label{Fig.2}
\end{figure}

Fig.~\ref{Fig.2} shows on a log-log plot the measured frequency
$f_{1}= n_{1} /N$ of single fluxon trapping, obtained by quenching
the sample $N$ times for each value of a given quenching time $\tau
_{Q}$, $n_{1}$ being the number of times that the inspection of the
low temperature JTJ current-voltage characteristics at the very end
of each thermal cycle showed that one defect was spontaneously
produced. $N$ ranged between $100$ and $2600$ and $n_{1}$ was never
smaller then $10$, except for the rightmost point for which
$n_{1}=3$ (and $N=1800$). The sample has undergone a total of more
than 100,000 thermal cycles without any measurable change of its
electrical parameters. The vertical error bars gives the statistical
error $f_{1}/\surd n_1$. The measurement of $\tau_Q$ by fitting
solutions to the heat equation follows that of \cite{Monaco1&2} in
its high accuracy. The relative error bars in $\tau _{Q}$ amounting
to $\pm10\%$ are as large as the dot's width.

\noindent To test Eq.(\ref{P1}), we have fitted the data with an
allometric function $f_{1}=a\,\tau _{Q}^{-b}$, with $a$ and $b$ as
free fitting parameters. A linear regression of $\log f_{1}$ vs.
$\log \tau _{Q}$, represented by the continuous line in
Fig.~\ref{Fig.2}, yields $b=0.51\pm 6\%$, in excellent agreement
with the predicted value $0.5$. The same fit gives $a=0.0092\pm
10\%$ (taking $\tau _{Q}$ in seconds) that is 6-7 times larger than
the predicted value $C/\sqrt{\xi_{0} \bar{c}_{nn}}$ (with $C=500 \mu
m$, $\xi _{0}=17 \mu m$ and $\bar{c}_{nn}=6 \cdot 10^3 m/s$). As a
bound we only expect agreement in the overall normalization $a$ to
somewhat better than an order of magnitude. Empirically, the
different condensed matter experiments have shown that the ratio
$a_{\rm observed}/a_{\rm predicted}$ varies widely from system to
system; $O(1)$ for superfluid $^3He$ \cite{grenoble, helsinki}, very
small for high-$T_c$ superconductors \cite{technion2}. We point out
that in our case the value of the prefactor $a$ is dependent,
although weakly, on the choice of $\alpha'$ and $f$, being
$C/\sqrt{\xi_{0} \bar{c}_{nn}} \propto (\alpha' /f)^{1/4}$. The
choice $f= 10 kHz$ was determined assuming that $1/a^2 \simeq 12
kHz$ is the characteristic frequency of our system at the time of
the thermal quench. As we noted earlier, the new scaling exponent
$\sigma = 0.5$, obtained by applying causality arguments to
\emph{realistic} JTJs, is twice as large than that observed earlier
\cite{Monaco1&2}, also in samples made with the same
$Nb-Al/Al_{ox}/Nb$ technology. The reason for this discrepancy
resides in the fact that at that time we were unaware of the high
sensitivity of $f_1$ to external magnetic fields which, although
small, were most likely present. In the present experiment much care
has been devoted both to avoid magnetic or current carrying
materials in the cryoprobe during the quench and to screen the chip
environment from $50Hz$ noise as well as from DC magnetic field.

The data of Fig.~\ref{Fig.2} resolve another issue. It has been
disputed that the Swihart velocity, which gives the behavior above,
is the relevant velocity for field ordering along the JTJ oxide. If
we had taken  the relevant velocity to be that of phase ordering in
the individual superconductors, as invoked by Zurek
\cite{zurek1,zurek2} when considering the spontaneous flux produced
on quenching annuli of simple superconductors, at the same level of
approximation we would have predicted $b = 0.25$ and the prefactor
orders of magnitude larger.

We have not observed the change in behavior predicting  $f_1$ to
change from the linear behavior with $C/{\bar\xi}$ of
Eq.(\ref{P1}) to a random walk behavior in the phase,

\begin{equation}
f_{1}\simeq \frac{C}{\bar{\xi}}=\frac{C}{\xi _{0}}\bigg(\frac{\tau _{Q}}{%
\tau _{0}}\bigg)^{-\sigma/2 },  \label{P2}
\end{equation}

\noindent once $f_1$ is sufficiently large. A first guess would
suggest that the transition from Eq.(\ref{P1}) to Eq.(\ref{P2})
would occur when $f_1\approx 1$. However, future experiments to be
carried out on ring-shaped JTJS having the circumference larger
than that of the sample reported in this paper and with $\tau
_{Q}$ in the same range of this experiment should clearly reveal
the transition from Eq.(\ref{P1}) to Eq.(\ref{P2}).

\medskip In summary, we see this experiment as providing unambiguous
corroboration of Zurek-Kibble scaling over a wide range of quenching
time $\tau_Q$ in accord with our predictions for $Nb-Al/Al_{ox}/Nb$
JTJs. As such, it replaces the experiment reported in
Ref.\cite{Monaco1&2} by being more realistic theoretically and more
sophisticated experimentally. We stress that this experiment is the
\textit{only} one to date to have confirmed the Zurek-Kibble scaling
exponent for a condensed matter system. Further experiments can be
devised to investigate the transition to the random walk regime and
the effect of the thermal gradients. Such experiments are in the
process of being performed.

\medskip The authors thank P. Dmitriev, A. Sobolev and M. Torgashin for the
sample fabrication and testing and U.L.Olsen for the help at the
initial stage of the experiment. This work is, in part, supported
by the COSLAB programme of the European Science Foundation, the
Danish Research Council, the Hartmann Foundation, the RFBR project
06-02-17206, and the Grant for Leading Scientific School
7812.2006.2.


\begin{references}
\bibitem{zurek1}  W.H. Zurek, {\it Nature} {\bf 317}, 505 (1985), {\it Acta
Physica Polonica} {\bf B24}, 1301 (1993).

\bibitem{zurek2}  W.H. Zurek, {\it Physics Reports} {\bf 276}, Number 4,
Nov. 1996.

\bibitem{kibble1}  T.W.B. Kibble, in {\it Common Trends in Particle and
Condensed Matter Physics}, {\it Physics Reports} {\bf 67}, 183
(1980).

\bibitem{KMR&MRK}  E. Kavoussanaki, R. Monaco and R.J. Rivers, {\it Phys. Rev.
Lett.} \thinspace {\bf 85}, 3452 (2000).  R. Monaco, R.J. Rivers
and E. Kavoussanaki, {\it Journal of Low Temperature
Physics}\thinspace {\bf 124}, 85 (2001).

\bibitem{Swihart}  J.C. Swihart, {\it J. Appl. Phys.,} {\bf 32}, 461 (1961).

\bibitem{Barone}  A. Barone and G. Paterno', {\it Physics and Applications
of the Josephson Effect, }John Wiley \& Sons, New York (1982).


\bibitem{Monaco1&2}  R. Monaco, J. Mygind, and R. J. Rivers, Phys. Rev. Lett.
89, 080603 (2002).  R. Monaco, J.Mygind, and R. J. Rivers,Phys.
Rev. B67, 104506 (2003).

\bibitem{grenoble}  C. Bauerle {\it et al.}, {\it Nature}\thinspace {\bf 382}%
, 332 (1996).

\bibitem{helsinki}  V.M.H. Ruutu {\it et al.}, {\it Nature}\thinspace {\bf %
382}, 334 (1996).

\bibitem{lancaster}  P.C. Hendry {\it et al}, {\it Nature}\thinspace {\bf 368%
}, 315 (1994).

\bibitem{lancaster2}  M.E. Dodd {\it et al.}, {\it Phys. Rev. Lett.}%
\thinspace {\bf 81}, 3703 (1998), {\it J. Low Temp. Physics}\thinspace {\bf %
15}, 89 (1999).

\bibitem{technion}  R. Carmi and E. Polturak, {\it Phys. Rev.} {\bf %
B\thinspace 60}, 7595 (1999).

\bibitem{technion2}  A. Maniv, E. Polturak, G. Koren, Phys. Rev. Lett. 91,
197001 (2003).

\bibitem{carmi}  R. Carmi, E. Polturak, and G. Koren, {\it Phys. Rev. Lett.}
{\bf 84}, 4966 (2000).

\bibitem{Kirtley}  J.R. Kirtley, C. C. Tsuei, and F. Tafuri, Phys. Rev.
Lett. 90, 257001 (2003).

\bibitem{pamplona}  S. Casado, W. Gonz\'{a}lez-Vi\~{n}as, H. Mancini and S.
Boccaletti, {\it Phys. Rev.\thinspace }{\bf E63}, 057301 (2001).

\bibitem{florence}  S. Ducci, P.L. Ramazza, W. Gonz\'{a}lez-Vi\~{n}as, and
F.T. Arecchi, {\it Phys. Rev. Lett.\thinspace }{\bf 83}, 5210
(1999).

\bibitem{AB} V. Ambegaokar and A. Baratoff, {\it Phys. Rev. Lett.}, {\bf 10}, 486
(1963). Errata, Phys. Rev. Lett., {\bf 11}, 104 (1963).

\bibitem{Rowell&Smith}  N.L. Rowell and H.J.T. Smith, {\it Can. J. Phys.,} {\bf 54}, 223 (1976).

\bibitem{golubov}  A.A. Golubov et al., {\it Phys. Rev. B,} {\bf 51}, 1073 (1995).

\bibitem{note1} This value comes from the Swihart formula
$\bar{c}_0=\sqrt {d_{ox}/{2 \lambda_{L0} \epsilon \mu}},$
with a $+20\%$ correction due to the presence of the idle region.

\bibitem{Next}  R. Monaco, J. Mygind,  M.\ Aaroe, V.P. Koshelets and R.J.
Rivers, in preparation.

\bibitem{VPK}  P.N. Dmitriev, I.L. Lapitskaya, L.V. Filippenko, A.B. Ermakov,
 S.V. Shitov, G.V. Prokopenko, S.A. Kovtonyuk, and V.P. Koshelets, {\it IEEE Trans. Appl.
Supercond} {\bf 13}, 107-110 (2003).

\end{references}
\end{document}